# Simultaneous Measurement of Thermal Conductivity and Heat Capacity Across Diverse Materials Using the Square-Pulsed Source (SPS) Technique

Tao Chen[1], Shangzhi Song[1], Yang Shen[1], Kexin Zhang[1], Puqing Jiang[1,*]

[1]*School of Power and Energy Engineering, Huazhong University of Science and Technology, Wuhan, Hubei 430074, China*

**ABSTRACT:** State-of-the-art techniques like dual-frequency Time-Domain Thermoreflectance (TDTR) and Frequency-Domain Thermoreflectance (FDTR) offer superb capability for simultaneous measurements of thermal conductivity and heat capacity with a spatial resolution on the order of 10 $\mu$m. However, their applicability is limited to highly conductive materials with an in-plane thermal conductivity exceeding 10 W/(m·K). In this paper, we introduce the Square-Pulsed Source (SPS) technique, offering a novel approach to concurrently measure thermal conductivity and heat capacity with a 10 $\mu$m spatial resolution, while significantly extending the measurable thermal conductivity range to an unprecedented low of 0.1 W/(m·K), offering enhanced versatility. To demonstrate and validate its efficacy, we conducted measurements on various standard materials—PMMA, silica, sapphire, silicon, and diamond—spanning a wide thermal conductivity range from 0.1 to 2000 W/(m·K). The obtained results exhibit remarkable agreement with literature values, with a typical measurement uncertainty of less than 10% across the entire thermal conductivity range. By providing a unique capability to characterize both highly and lowly conductive materials with micron-scale spatial resolution, the SPS method opens new avenues for understanding and engineering thermal properties across diverse applications.

*Corresponding Author: jpq2021@hust.edu.cn

**KEYWORDS:** Thermal conductivity, Heat capacity, Square-pulsed source (SPS) method, Thermoreflection, micro-scale spatial resolution

## 1. INTRODUCTION

The measurement of thermal conductivity ($k$) and heat capacity ($C$) holds paramount importance across various scientific and industrial applications [1, 2]. Thermal conductivity quantifies a material's ability to conduct heat, providing indispensable insights for applications such as thermal management in electronics [3], architectural insulation [4], and the development of energy-efficient materials [5]. Similarly, understanding heat capacity is indispensable, as it dictates the extent to which a material's temperature will change in response to heat absorption or release, thereby exerting a profound impact on various applications ranging from thermal energy storage to refrigeration systems [6, 7]. Accurate measurement of these thermal properties greatly advances renewable energy [8], climate control [9], and energy efficiency [10], ultimately shaping the landscape of modern engineering and technology.

The process of gauging $k$ and $C$ inherently involves heat-related phenomena, as thermal properties manifest during such processes. Measurement methods vary based on applied heat and temperature measurement techniques. Traditional approaches for measurement of $k$ include steady-state methods [11, 12], hot wire methods [13], laser flash methods [14], and planar heat source methods [15]. Likewise, conventional methods for measuring $C$ encompass calorimetric techniques such as differential scanning calorimetry [16], adiabatic calorimetry [17], and isothermal calorimetry [18]. However, these conventional measurement techniques often face limitations, particularly their reliance on larger sample sizes in the millimeter or centimeter range, making them less suitable for sub-millimeter-sized samples.

In recent years, microscale thermal property measurement techniques grounded in laser thermoreflectance have gained increasing attention. Their appeal lies in minimal sample size requirements and high measurement accuracy, providing a promising avenue for advancing the precision of thermal property measurements in smaller-scale materials [19, 20]. Dual-frequency time-domain thermoreflectance (TDTR) offers simultaneous measurement of $k$ and $C$ with micron-scale spatial resolution [21, 22]. This is achieved by using high-frequency (e.g., 10 MHz) measurements to probe the sample's cross-plane thermal effusivity ($\sqrt{k_z C}$) and low-frequency (e.g., 0.1 MHz) measurements to probe the sample's in-plane thermal diffusivity ($k_r/C$). If the sample is isotropic with $k_r = k_z = k$, both the thermal conductivity and heat capacity can be determined from the measured $\sqrt{k_z C}$ and $k_r/C$. However, TDTR can only be applied to materials with high in-plane thermal conductivity ($k_r > 10 \text{ W/(m·K)}$), mainly limited by its lowest measurable frequency of 0.1 MHz [23, 24]. An alternative technique, frequency-domain thermoreflectance (FDTR), shares the same limitation on the lowest measurable in-plane thermal conductivity [25, 26].

In this context, our paper introduces a groundbreaking methodology—the Square-Pulsed Source (SPS) technique—that transcends the constraints of existing techniques. By combining features of both TDTR and FDTR, the SPS technique retains the capability for concurrent $k$ and $C$ measurements with micrometer-scale spatial resolution. Notably, the SPS technique allows for a significantly expanded modulation frequency range from 10 MHz down to 1 Hz, thereby broadening the measurable in-plane $k$ range from 2000 W/(m·K) down to an unprecedented low of 0.1 W/(m·K). This breakthrough is not just a technical achievement but a response to the growing demand for comprehensive thermal characterization, particularly in low-conductivity materials.

In the subsequent sections of this paper, we delve into the development and validation of the SPS technique, showcasing its efficacy in simultaneously measuring $k$ and $C$ across a range of standard materials. Through this exploration, we aim to highlight not only the technical prowess of the SPS method but also its transformative potential for advancing our understanding and manipulation of thermal properties in materials across diverse scientific and engineering applications.

## 2. METHODOLOGIES

### 2.1. Principle of the SPS technique

The operational principle of the SPS technique is depicted in Fig. 1(a). This technique utilizes a modulated pump laser employing a 50% duty cycle square wave to periodically heat the sample surface. Simultaneously, another laser is used to capture the temperature response of the sample surface through the principle of thermoreflectance. The acquired temperature response signals undergo normalization in both amplitude and time axes. Subsequently, a heat transfer model is applied for optimal fitting of the normalized signals, facilitating the extraction of the desired thermal properties of the specimen.

The success of the SPS technique relies on three pivotal factors: firstly, achieving high-speed signal acquisition with a high signal-to-noise ratio spanning a broad frequency range; secondly, ensuring rapid and accurate simulation of measurement signals; and thirdly, judiciously selecting square wave modulation frequencies and laser spot sizes for measurements to enhance signal sensitivity to the desired thermal properties, thereby minimizing measurement errors. The three factors correspond to three aspects of the technique respectively, i.e., the optimized experimental setup, the accurate thermal modeling, and the optimal parameter-selection rules, which will be elaborated below.

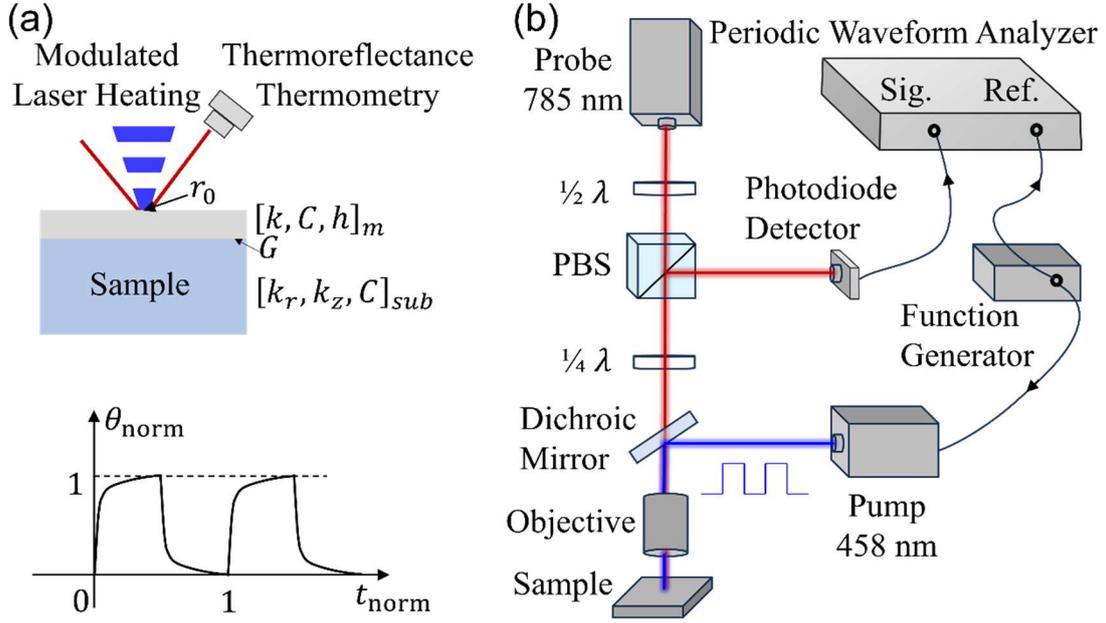

**Fig. 1** (a)Schematic of the measurement principle of SPS and sample structure. (b) Schematic diagram of the SPS setup.

*2.2. Experimental setup*

A schematic diagram of the experimental setup of SPS is shown in Fig. 1(b). A function generator generates a square wave function at a preset frequency, which is transmitted to modulate the output of the pump laser. The modulated pump laser, directed by a dichroic mirror, enters the objective lens and focuses on the sample surface, inducing periodic heating. Simultaneously, another laser with a different wavelength passes through the dichroic mirror, entering the same objective lens to focus on the sample surface, coinciding with the pump laser spot. Based on the principle of thermoreflectance, the surface reflectance of the sample is linearly correlated with the temperature change if not exceeding 10 K. Consequently, the detected light reflected from the sample surface carries information about the temperature variations, captured by a photodetector. The photodetector converts the optical signal into an electrical signal, collected by a high-speed analog-to-digital converter operating at 14 bits and 2.6 Gsps. Substantial raw data is then transmitted to the graphics processing unit (GPU), where parallel computing architecture facilitates rapid processing, swiftly yielding a high signal-to-noise ratio temperature amplitude signal of a single heating cycle, completing the signal acquisition.

In this system, the key to achieving rapid acquisition of high signal-to-noise ratio signals across a wide frequency range from 1 Hz to 10 MHz lies in the high-speed data acquisition module and the parallel computing architecture of the GPU. The bandwidths of both the photodetector and the data acquisition card exceed 1 GHz, favorably accommodating measurements at the high frequency of 10 MHz. An alternative and convenient option is to use the Periodic Waveform Analyzer (PWA) of Zurich lock-in amplifier UHFLI, which has a bandwidth of 600 MHz. When acquiring signals at low modulation frequencies (<100 kHz), using a balanced photodiode detector can substantially reduce the $1/f$ noise.

During the thermoreflectance measurement, the sample is typically coated with a thin metal layer of around 100 nm thick as the transducer. The wavelength of the laser can be flexibly selected according to the experiment's requirements, with common choices for the probe laser being 532 nm or 785 nm. For a probe laser wavelength of 532 nm, materials such as Au or Ta are suitable for the metal transducer layer. Alternatively, for a probe laser wavelength of 785 nm, materials like Al or Ta can be employed. These metal materials exhibit high thermoreflectance coefficients at their respective wavelengths, enhancing the intensity of the measurement signal.

*2.3. Thermal model*

For $k$ and $C$ measurements in a multilayered configuration, the analytical solution of the heat diffusion equation in the frequency domain emerges as a notably efficient approach. Hankel and Fourier transformations adeptly transform the inherently complex three-dimensional heat diffusion equation in the time domain into a streamlined one-dimensional query in the frequency domain. This elegant transformation allows for the direct application of heat flux on the surface of the first layer, which then methodically propagates downward. Within this framework, the correlation between the temperature and heat flux density at the bottom of the last layer ($\Theta_{b,n}$, $Q_{b,n}$) with those on the surface of the first layer ($\Theta_{t,1}$, $Q_{t,1}$) is elegantly encapsulated through an array of transmission matrices [19]:

$$\begin{bmatrix} \Theta_b \\ Q_b \end{bmatrix}_n = M_n R_{n-1} \ldots M_2 R_1 M_1 \begin{bmatrix} \Theta_t \\ Q_t \end{bmatrix}_1 = \begin{bmatrix} A & B \\ C & D \end{bmatrix} \begin{bmatrix} \Theta_t \\ Q_t \end{bmatrix}_1 \tag{1}$$

Here, $M_i = \begin{pmatrix} \cosh \lambda l & -\frac{1}{k_z \lambda} \sinh \lambda l \\ -k_z \lambda \sinh \lambda l & \cosh \lambda l \end{pmatrix}_i$, and $R_i = \begin{pmatrix} 1 & -1/G \\ 0 & 1 \end{pmatrix}_i$, where $\lambda_i = \sqrt{4\pi^2 \rho^2 \eta + i\omega C/k_z}$, with $\rho$ and $\omega$ representing the frequency domain variables for Hankel and Fourier transformations, respectively, $\eta = k_r/k_z$, $l_i$ specifies the thickness of the $i$-th layer, and $G_i$ is the thermal conductance across the interface under the $i$-th layer.

Under the assumption that the last layer is semi-infinite, its bottom heat flux density, $Q_{b,n}$, effectively becomes zero, i.e., $Q_{b,n} = C\Theta_{t,1} + DQ_{t,1} = 0$. This assumption facilitates the derivation of the temperature response function as:

$$\hat{G} = \frac{\Theta_0}{Q_0} = \frac{\Theta_{t,1}}{Q_{t,1}} = -\frac{D}{C} \tag{2}$$

Here, $Q_0$ signifies the applied heat flux density, with $\Theta_0$ representing the detected temperature response.

The applied heat flux density through the modulated pump laser is characterized by a Gaussian spatial profile with a $1/e^2$ radius of $r_1$, and temporally as a square wave function at frequency $f_0$. Through the application of Hankel and Fourier transformations, the frequency domain representation of the applied heat flux $Q_0$ is derived as follows:

$$Q_0(\rho, \omega) = A_0 \exp\left(-\frac{\pi^2 \rho^2 r_1^2}{2}\right) \left(\frac{\delta(\omega)}{2} + \frac{1}{\pi} \sum_{n=1}^{N} i \frac{\left(\delta(\omega + 2\pi(2n-1)f_0) - \delta(\omega - 2\pi(2n-1)f_0)\right)}{2n-1}\right) \quad (3)$$

Here, $A_0$ is the amplitude of the absorbed pump power, $\delta(x)$ is the Dirac delta function, and $N$ represents the maximum number of harmonics captured for the square wave function, determined as $[f_{bw}/f_0]$, with $f_{bw}$ chosen as the smaller value between the bandwidths of the photodiode detector and the data acquisition card.

Substituting this expression into Eq. (2), the frequency domain temperature response $\Theta_0$ can be ascertained. Inverting Hankel and Fourier transformations of $\Theta_0$ yields the temporal temperature change $\Delta T_t$, which, after spatial weighting via a Gaussian distribution by a probe laser with a $1/e^2$ radius of $r_2$, culminates in the detector-measured temporal signal variation as:

$$\Delta T(t) = \frac{A_0}{2} \int_{-\infty}^{\infty} \hat{G}(\rho, 0) \exp(-\pi^2 \rho^2 r_0^2) \, 2\pi\rho \mathrm{d}\rho$$
$$-2A_0 \mathrm{Re}\left\{\sum_{n=1}^{N} \frac{i e^{i2\pi(2n-1)f_0 t}}{(2n-1)\pi} \int_{-\infty}^{\infty} \hat{G}(\rho, 2\pi(2n-1)f_0) \exp(-\pi^2 \rho^2 r^2) \, 2\pi\rho \mathrm{d}\rho\right\} \quad (4)$$

where $r_0 = \sqrt{(r_1^2 + r_2^2)/2}$ is the average spot size of the pump and probe, $i = \sqrt{-1}$ is the imaginary unit, and $\mathrm{Re}\{z\}$ represents taking the real part of the complex number $z$.

$\Delta T(t)$ can be calculated numerically. After further normalization and comparison against experimental signals, the thermal properties can be extracted. The thermal model in the SPS technique shares the same root with the well-developed FDTR [27], differing primarily in the expression for the heat flux $Q_0$, thereby ensuring great reliability for precision thermal measurements.

*2.4. Heat conduction regimes and the principle of concurrent k and C measurement*

In SPS measurements, the heat conduction regimes are jointly determined by the laser spot size and modulation frequency, influencing the signal sensitivity to different material properties. Specifically, as shown in Figure 2, the comparison between the in-plane thermal diffusion length (expressed as $d_{f,r} = \sqrt{k_r/\pi f_0 C}$) and the laser spot diameter $2r_0$ governs the conduction mode during measurement.

When measured using a large laser spot and a low enough modulation frequency such that $d_{f,r} \approx 2r_0$, heat primarily conducts radially. In this scenario, the measured signal is predominantly influenced by the ratio between $d_{f,r}$ and $2r_0$, denoted as the combined parameter $k_r/Cr_0^2$. Therefore, with the laser spot size $r_0$ predetermined, the in-plane thermal diffusivity of the sample $k_r/C$ can be accurately determined. Practically, this condition can be more flexible, with the signals primarily sensitive to $k_r/Cr_0^2$ as long as $\frac{1}{3}r_0 < d_{p,r} < 8r_0$.

When measured using a large laser spot and a high modulation frequency such that $d_{f,r}$ is much smaller than $\frac{1}{3}r_0$, heat mainly transfers one-dimensionally in the through-plane direction. In this case, the measured signal is mainly affected by the cross-plane thermal effusivity $\sqrt{k_z C}$ and the metal film's specific heat capacity per unit area $h_m C_m$, approximately $\frac{\sqrt{k_z C}}{h_m C_m}$. Here, the subscript $m$ represents the metal film's properties, with $h_m$ being the thickness of the metal film. At this time, the signal is also sensitive to the thermal conductance $G$ at the interface between the metal film and the sample. Therefore, when the parameter to be measured is the material's cross-plane thermal conductivity $k_z$, an appropriate laser spot and modulation frequency should be selected such that $d_{p,r}$ and $2r_0$ fall within this range.

For isotropic materials with $k_r = k_z = k$, the thermal conductivity $k$ and heat capacity $C$ can be simultaneously derived from the measured in-plane thermal diffusivity ($k_r/C$) and cross-plane thermal effusivity ($\sqrt{k_z C}$) from the above two sets of measurements.

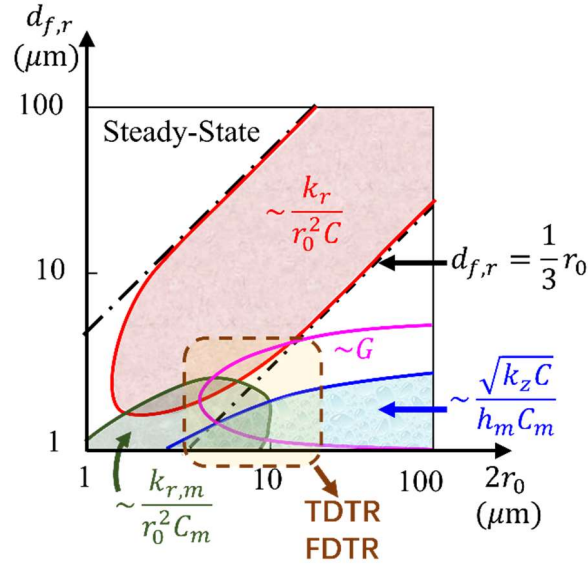

**Fig. 2** Different heat conduction regimes achievable in SPS experiments, determined by the relative lengths of laser spot diameter $2r_0$ and the in-plane thermal diffusion length $d_{f,r}$. Here, $d_{f,r}$ is defined as $d_{f,r} = \sqrt{k_r/\pi f_0 C}$, with $f_0$ being the modulation frequency. Comparatively, the capability of TDTR and FDTR in probing the combined parameter of $k_r/Cr_0^2$ is significantly restricted due to their limited modulation frequency range.

The SPS technique can also be applied to measure the in-plane thermal diffusivity of the metal transducer layer, $k_{r,m}/C_m$, if measured using a small laser spot and a high modulation frequency such that $d_{f,r} < r_0 < 5\ \mu m$, where the signal is predominantly sensitive to $\frac{k_{r,m}}{r_0^2 C_m}$. This sensitivity can be further enhanced if the substrate is a material with low thermal conductivity, as it facilitates confining the heat within the metal transducer layer.

The regime with a small laser spot and a low modulation frequency such that $d_{p,r} \geq 8r_0$ should be avoided in SPS experiments. In such cases, the temperature rise quickly reaches a steady state. Consequently, the measured signal, once normalized, may lose its relevance for fitting.

A notable advantage of the SPS technique over other similar thermorefletance techniques is its significantly broad range of achievable modulation frequencies, spanning from 1 Hz to 10 MHz. In comparison, techniques like TDTR and FDTR have limited modulation frequency ranges, consequently restricting the measurable in-plane thermal conductivity to values larger than $10 \text{ W}/(\text{m} \cdot \text{K})$.

## 3. RESULTS AND DISCUSSION

### 3.1. Sample preparation

To demonstrate the versatility of SPS in the simultaneous measurement of $k$ and $C$ across a diverse range of materials, five samples were selected, featuring thermal conductivities spanning four orders of magnitude, from 0.1 to $2000 \text{ W}/(\text{m} \cdot \text{K})$. These samples include polymethyl methacrylate (PMMA), fused silica, sapphire, single-crystal silicon, and single-crystal CVD diamond, with their nominal $k$ and $C$ values listed in Table 1. Of these samples, PMMA (optical grade) was provided by Kunxin Co. Ltd (www.chinaacrylicsheet.com), while fused silica, sapphire, and single-crystal silicon wafers were purchased from MTI. Additionally, the single-crystal CVD diamond sample was generously provided by Prof. Junjun Wei from Beijing University of Science and Technology.

To conduct the measurements, all samples were coated with an approximately 100 nm thick layer of aluminum using magnetron sputtering. The thickness of this metal transducer layer was determined using a step profiler. Subsequently, the thermal conductivity of the metal film was calculated using the Wiedemann-Franz law, based on the electrical resistivity measured by the van der Pauw method. The volumetric heat capacity of the transducer layer was obtained from relevant literature databases [28]. The size of the laser spot used in the experiments was determined by the knife-edge method.

In what follows, three specific measurement examples on PMMA, Si, and diamond are presented to demonstrate the SPS method's capability to simultaneously measure thermal conductivity and heat capacity across a broad range.

*3.2. PMMA*

PMMA demonstrates a significantly low thermal conductivity, which poses a notable challenge for traditional measurement techniques such as TDTR and FDTR. These methods are constrained by their ability to accurately measure $k_r$ only above 10 W/(m · K). Within this framework, the distinctive advantages of the SPS method, as developed in this study, become apparent. Characterized by its versatility and heightened sensitivity, the SPS method offers a robust solution for precisely measuring materials with a low $k_r$.

Figure 3 presents the measurement outcomes for PMMA. Employing a moderate laser spot size of 11.5 $\mu$m and a low modulation frequency of 50 Hz for the measurement, the signals are exclusively sensitive to the in-plane thermal diffusivity ($k_r/C$) of PMMA and the laser spot size $r_0$ (Fig. 3(a1 − c1)). In Fig. 3(a1), the measured signals (symbols) are compared with the thermal model prediction (curve) for a full heating period, displayed on a linear scale. To closely scrutinize signal sensitivity and fitting quality, signals within the normalized time range of 0.55 to 0.7 are plotted in a log-log scale in Fig. 3(b1). Predicted signals with 30% bounds of the best-fitted $k_r$ value are added as dashed curves as a guide for the sensitivity of the signals to $k_r$ of PMMA. The sensitivity of the signal $A_{\text{norm}}$ to an arbitrary parameter, $\xi$, can be quantified by defining the sensitivity coefficient as $S_\xi = \partial \ln A_{\text{norm}} / \partial \ln \xi$. Figure 3 (c1) shows the sensitivity coefficients of the $A_{\text{norm}}$ signals from Fig. 3(b1) to all the thermal system parameters, showcasing exclusively to parameters $k_r$, $C$, and $r_0$. Notably, sensitivity to $k_r$ exhibits the same magnitude and opposite sign compared to sensitivity to $C$, while sensitivity to $r_0$ is twice as that to $C$, suggesting a sensitivity to the combined parameter $k_r/Cr_0^2$. With $r_0$ predetermined, PMMA's in-plane thermal diffusivity $k_r/C$ can be accurately determined.

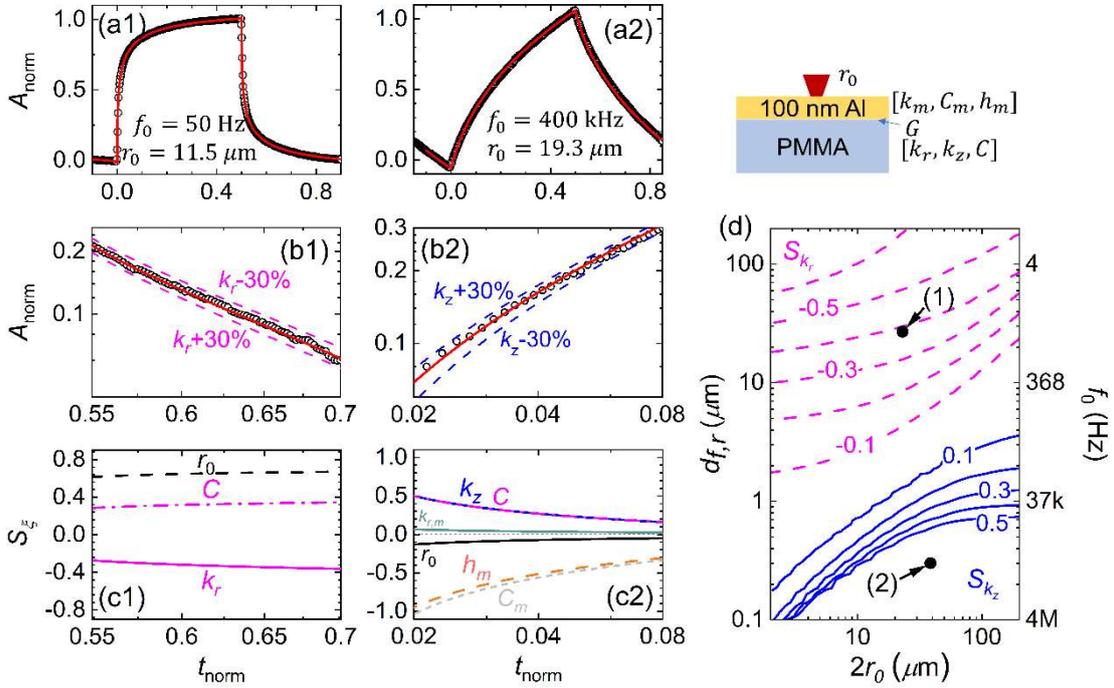

**Fig.3** (a, b): SPS signals for PMMA obtained with different configurations of (a1, b1) with $r_0 = 11.5\ \mu m$ and $f_0 = 50$ Hz and (a2, b2) with $r_0 = 19.3\ \mu m$ and $f_0 = 400$ kHz, compared to the thermal model predictions. (c): Sensitivity coefficients ($S_\xi$) of the SPS signal shown in (b) concerning all parameters in the thermal model, where $S_\xi$ is defined as $S_\xi = \partial \ln A_{\text{norm}} / \partial \ln \xi$. (d): Contours of sensitivity coefficients for the thermal conductivity in the z-direction ($k_z$, depicted by black solid curves) and the radial direction ($k_r$, shown as red dashed curves) plotted against spot diameter ($2r_0$) and thermal diffusion length ($d_{f,r}$). Here, $d_{f,r}$ is defined as $d_{f,r} = \sqrt{k_r/\pi f_0 C}$. The experimental configurations for the two sets of measurements are indicated by black dots on the plots.

Conversely, employing a larger laser spot size of 19.3 $\mu$m and a relatively high modulation frequency of 400 kHz for the measurement, the signals become exclusively sensitive to the cross-plane thermal effusivity ($\sqrt{k_z C}$) of PMMA and the heat capacitance ($h_m C_m$) of the transducer layer, as depicted in Fig. 3(a2 – c2). In Fig. 3(a2), the measured signals (symbols) are compared with the thermal model prediction (curve) for a full heating period, plotted on a linear scale. Signals within the normalized time range of 0.02 to 0.08 are plotted in a log-log scale in Fig. 3(b2), with the predicted signals having 30% bounds of the best-fitted $k_z$ value added as dashed curves to guide the sensitivity of the signals to $k_z$ of PMMA. The sensitivity plot in Fig. 3(c2) shows that this group of signals is exclusively sensitive to parameters $k_z$, $C$, $h_m$ and $C_m$, with sensitivity coefficients to $h_m$ and $C_m$ having an opposite sign and a magnitude twice as that to $k_z$ and $C$, suggesting a sensitivity to the combined parameter $\sqrt{k_z C}/h_m C_m$. With $h_m$ and $C_m$ predetermined, PMMA's cross-plane thermal effusivity $\sqrt{k_z C}$ can be accurately determined.

By combining these two sets of measurements and applying the condition that PMMA is isotropic with $k = k_r = k_z$, the thermal conductivity of PMMA is determined as $0.19 \pm 0.013$ W/(m·K), and its heat capacity as $1.644 \pm 0.11$ MJ/(m$^3$·K), both with an uncertainty of ~7%. Please note that the entire dataset has been optimally fitted to extract these thermal properties, although only a small portion of data is displayed in Fig. 3(b) and (c) for visualization purposes.

Figure 3 (d) shows the contours of sensitivity coefficients for $k_z$ (depicted by black solid curves) and $k_r$ (represented by red dashed curves), plotted against $2r_0$ and $d_{f,r}$. The experimental configurations selected for the two sets of measurements are denoted by black dots in the plot. Notably, regions sensitive to $k_z$ are clearly distinguished from those sensitive to $k_r$, allowing for their independent determination. Specifically, signals exhibit predominant sensitivity to $k_r$ when measured with a low modulation frequency such that $d_{f,r} \approx 2r_0$, and to $k_z$ when measured with a high modulation frequency such that $d_{f,r}$ is considerably smaller than $r_0/3$. These observations align with the general guideline instructions outlined in Section 2.4.

*3.3. Si*

Single-crystal silicon exhibits a thermal conductivity three orders of magnitude higher than that of PMMA, nevertheless, its $k$ and $C$ can still be determined utilizing the same underlying principle. Figure 4 displays the measurement results for single-crystal silicon. Using a large laser spot size of 20 $\mu$m and a low modulation frequency of 20 kHz for the measurement, the signals are primarily sensitive to the in-plane thermal diffusivity ($k_r/C$) of single-crystal silicon and the laser spot size $r_0$ (Fig. 4(a1 – c1)). In Fig. 4(a1), the measured signals (symbols) are compared with the thermal model prediction (curve) over a complete heating cycle, displayed on a linear scale. To closely examine the signal sensitivity and fitting quality, signals within the normalized time range of 0.5 to 0.65 are plotted on a log-log scale in Fig. 4(b1). Predicted signals with 30% bounds of the best-fitted $k_r$ value are added as dashed curves, serving as a guide for the sensitivity to $k_r$ of single-crystal silicon. Fig. 4(c1) shows the sensitivity coefficients of the $A_{\text{norm}}$ signals from Fig. 4(b1) to all thermal system parameters, predominantly showcasing sensitivity to $k_r$, $C$, and $r_0$, indicating a primary sensitivity to the combined parameter $k_r/Cr_0^2$. With $r_0$ predetermined, the in-plane thermal diffusivity $k_r/C$ of single-crystal silicon can be accurately determined.

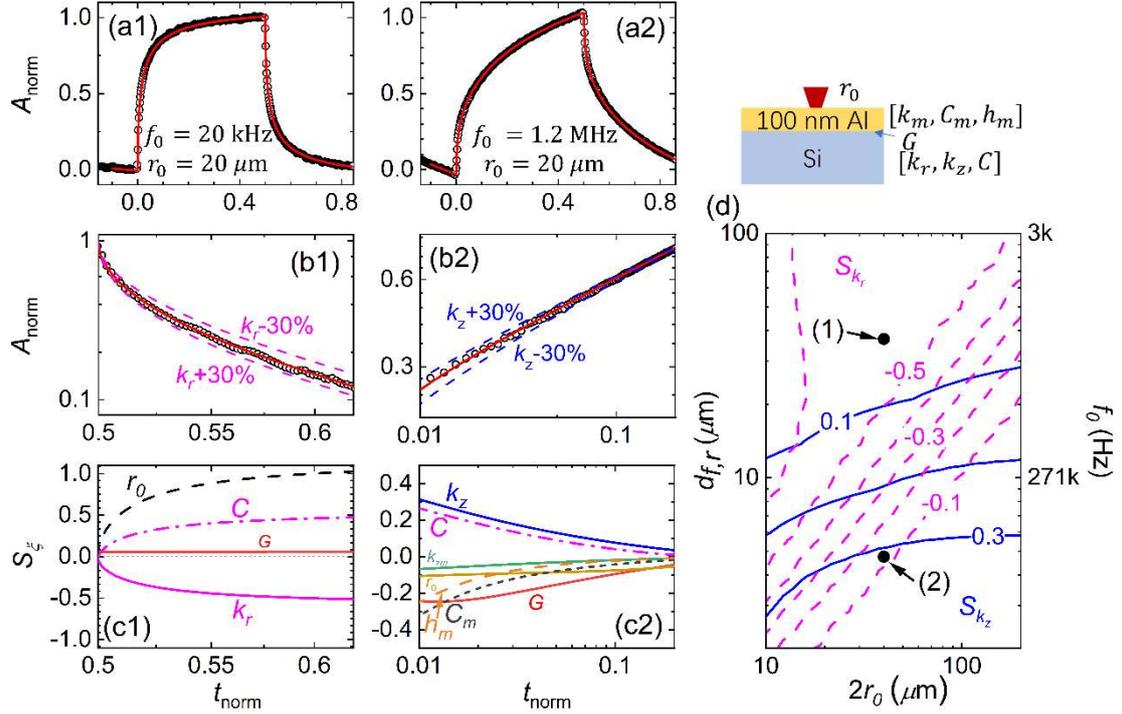

**Fig.4** (a, b): SPS signals for single-crystal Si obtained using a fixed spot size of 20 μm and different modulation frequencies of (a1, b1) $f_0 = 20$ kHz and (a2, b2) $f_0 = 1.2$ MHz, compared to the thermal model predictions. (c): Sensitivity coefficients ($S_\xi$) of the SPS signal shown in (b) concerning all parameters in the thermal model, where $S_\xi$ is defined as $S_\xi = \partial \ln A_{\text{norm}} / \partial \ln \xi$. (d): Contours of sensitivity coefficients for the thermal conductivity in the z-direction ($k_z$, depicted by black solid curves) and the radial direction ($k_r$, shown as red dashed curves) plotted against spot diameter ($2r_0$) and thermal diffusion length ($d_{f,r}$). Here, $d_{f,r}$ is defined as $d_{f,r} = \sqrt{k_r / \pi f_0 C}$. The experimental configurations for the two sets of measurements are indicated by black dots on the plots.

Conversely, using the same laser spot size of 20 μm and a relatively high modulation frequency of 1.2 MHz for the measurement, the signals are primarily sensitive to the cross-plane thermal effusivity ($\sqrt{k_z C}$) of the substrate, the heat capacitance ($h_m C_m$) of the transducer layer, and the interface thermal conductance $G$ between the transducer and the substrate, as depicted in Fig. 4(a2 – c2). In Fig. 4(a2), the measured signals (symbols) are compared with the thermal model prediction (curve) over a complete heating cycle, displayed on a linear scale. Signals within the normalized time range of 0.01 to 0.2 are plotted on a log-log scale in Fig. 4(b2). The sensitivity plot in Fig. 4(c2) reveals that this set of signals is mainly sensitive to the parameters $k_z$, $C$, $h_m$, $C_m$, and $G$. It is interesting to note that in this case, sensitivity coefficients to $h_m$ and $C_m$ still have an opposite sign compared to those associated with $k_z$ and $C$. However, their magnitudes are similar, unlike the scenario depicted in Fig. 3(c2) where the sensitivity coefficients to $h_m$ and $C_m$ are twice as large as those related to $k_z$ and $C$. This discrepancy arises from the relatively high thermal conductivity of the silicon substrate, which causes the signals to be sensitive not only to the same combined parameter of $\sqrt{k_z C}/h_m C_m$, but also to other parameters including $\sqrt{k_{z,m} C_m}/h_m C_m$, $C_m/k_{r,m}$, and $h_m/k_{z,m}$. The sensitivity to $h_m$ and $C_m$ shown in Fig. 4(c2) reflects a combined effect of sensitivity to all these pertinent parameters. The sensitivity to the interface thermal conductance $G$, on the other hand, is uncoupled from other parameters, allowing $G$ to be fitted independently as $150 \pm 11$ MW/(m² · K). With $h_m$ and $C_m$ as inputs, the cross-plane thermal effusivity $\sqrt{k_z C}$ of single-crystal silicon can be accurately determined.

By integrating these two sets of measurements and applying the condition that single-crystal silicon is isotropic with $k = k_r = k_z$, the thermal conductivity of single-crystal silicon is determined as $142 \pm 8$ W/(m · K), and its heat capacity is determined as $1.665 \pm 0.093$ MJ/(m³ · K), both with an uncertainty of 6%.

*3.4. Diamond*

The SPS method can also be utilized to measure materials with exceptionally high thermal conductivities, such as single-crystal diamonds, albeit with some additional considerations. One critical aspect is the selection of the laser spot size. If the spot size is too small, the temperature quickly stabilizes during the heating and cooling phases, making it hard to extract meaningful thermal properties. Conversely, opting for a spot size that is too large reduces the heat flux, leading to a decrease in the signal-to-noise ratio.

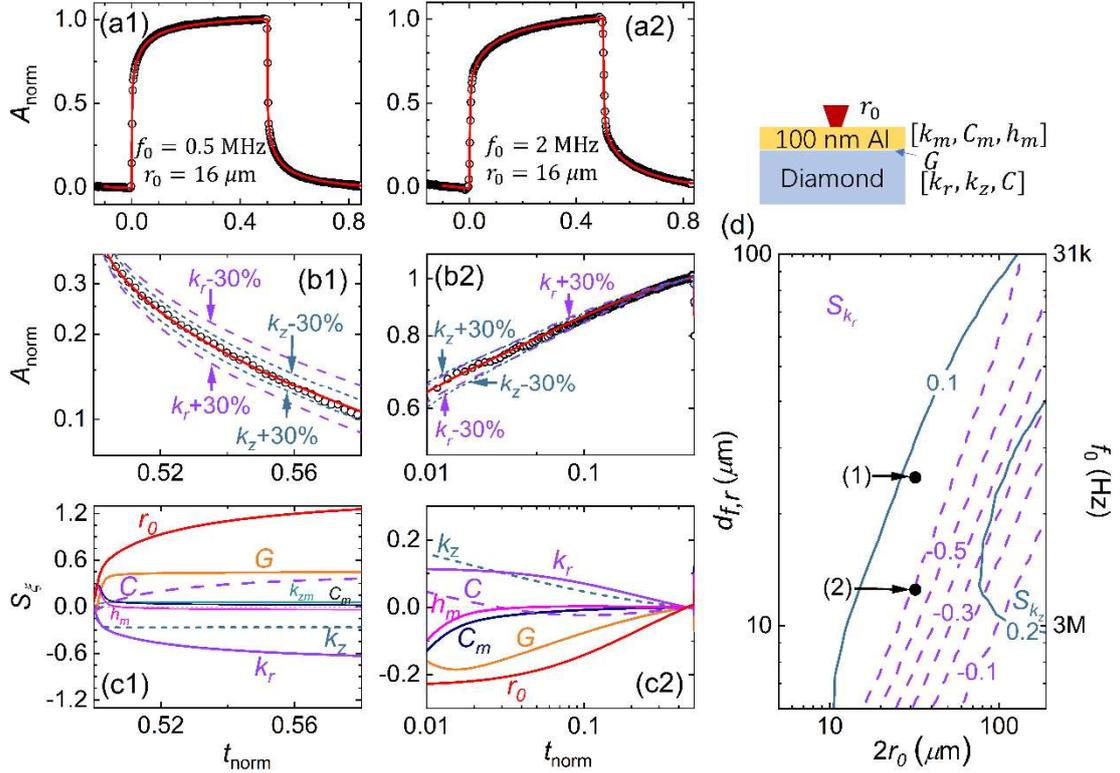

**Fig.5** (a, b): SPS signals for single-crystal diamond obtained using a fixed spot size of 16 $\mu$m and different modulation frequencies of (a1, b1) $f_0 = 500$ kHz and (a2, b2) $f_0 = 2$ MHz, compared to the thermal model predictions. (c): Sensitivity coefficients ($S_\xi$) of the SPS signal shown in (b) concerning all parameters in the thermal model, where $S_\xi$ is defined as $S_\xi = \partial \ln A_{\text{norm}} / \partial \ln \xi$. (d): Contours of sensitivity coefficients for the thermal conductivity in the z-direction ($k_z$, depicted by black solid curves) and the radial direction ($k_r$, shown as red dashed curves) plotted against spot diameter ($2r_0$) and thermal diffusion length ($d_{f,r}$). Here, $d_{f,r}$ is defined as $d_{f,r} = \sqrt{k_r/\pi f_0 C}$. The experimental configurations for the two sets of measurements are indicated by black dots on the plots.

Figure 5 shows our measurement results for the single-crystal diamond, employing a fixed laser spot size of 16 $\mu$m. When measured using a relatively low modulation frequency of 500 kHz (refer to Fig. 5(a1 – c1)), the signals are predominantly sensitive to $k_r/Cr_0^2$, with some additional sensitivity to $\sqrt{k_z C}$ and $G$, as shown in Fig. 5(c1). Conversely, when employing a higher modulation frequency of 2 MHz, the signals become sensitive to $\sqrt{k_z C}$, $k_r/Cr_0^2$, and $G$, simultaneously, as depicted in Fig. 5(c2). Given the differing degrees of sensitivity to $k_z C$ and $k_r/C$ in these two datasets, these two parameters can be simultaneously determined through best-fitting. By imposing the condition of isotropy for single-crystal diamonds with $k = k_r = k_z$, we determine the thermal conductivity of the single-crystal diamond as $1890 \pm 173$ W/(m · K), with an uncertainty of 9.2%, and its heat capacity as $1.93 \pm 0.08$ MJ/(m³ · K), with an uncertainty of 4%. Furthermore, $G$ remains decoupled from other parameters, particularly at higher frequencies, allowing its determination as $176 \pm 10$ MW/(m² · K).

The sensitivity contour shown in Fig. 5(d) reveals that the sensitivity for $k_r$ and $k_z$ of the diamond cannot be entirely isolated, which contrasts with the scenario observed for low-conductivity materials like PMMA. Nevertheless, it is always possible to find two different configurations where one is more sensitive to $k_r$ and the other to $k_z$. Through careful best-fitting of these datasets, the simultaneous determination of $k_r/C$ and $\sqrt{k_z C}$ is achievable. Additionally, Fig. 5(d) illustrates that the sensitivity to both $k_r$ and $k_z$ of diamond undergoes significant changes with variations in the laser spot size. Therefore, a slight increase in the laser spot size can notably suppress the sensitivity to $k_r$ while enhancing the sensitivity to $k_z$ of diamond. Nevertheless, we employed a fixed spot size of 16 $\mu$m for this measurement due to the limited pump power of 75 mW available for our system. However, this limitation should not be considered intrinsic to the SPS technique. If a laser with higher power is utilized for the system, a larger laser spot size can be employed, which is more favorable to measuring high-conductivity meatierlas like diamonds.

*3.5. Summary*

Figure 6 summarizes the thermal conductivity and heat capacity values obtained for PMMA, fused silica, sapphire, single-crystal silicon, and single-crystal diamond using the SPS method at room temperature, compared with the accepted literature values [29-38]. The percentage deviations of the measured values from the literature values are all less than 6%. The thermal conductivities studied here cover a broad range from 0.1 to 2000 W/(m · K). All data are also listed in Table I. Overall, this new SPS method demonstrates exceptional performance in the concurrent measurement of thermal conductivity and heat capacity for isotropic materials, showing excellent agreement with the literature values across the entire thermal conductivity range.

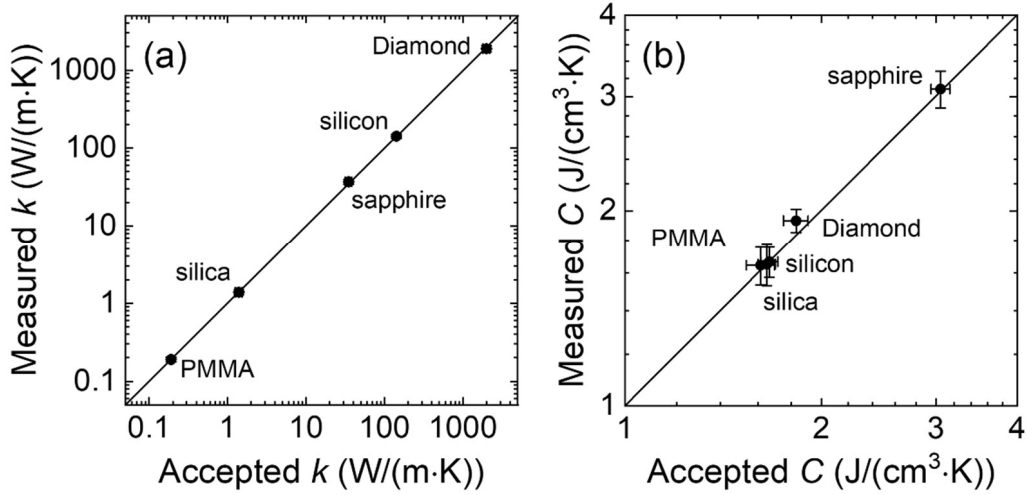

**Fig. 6** Measurement of the thermal conductivity (a) and heat capacity (b) of different reference samples at room temperature using the new SPS method, with comparisons to literature values for PMMA [29, 30], fused silica [31, 32], sapphire [33, 34], silicon [35, 36], and diamond [37, 38], respectively. Error bars are included for all data points; however, those in (a) are smaller than the symbol size and thus not visible.

**Table I.** The literature and measured values for thermal conductivity ($k$) and volumetric heat capacity ($C$) of different reference samples.

| Sample | Literature | | Current | |
|---|---|---|---|---|
| | $C$ (MJ/(m³ · K)) | $k$ (W/(m · K)) | $C$ (MJ/(m³ · K)) | $k$ (W/(m · K)) |
| PMMA | $1.614 \pm 0.08$ [29] | $0.19 \pm 0.004$ [30] | $1.644 \pm 0.11$ | $0.19 \pm 0.01$ |
| Fused silica | $1.65 \pm 0.05$ [31] | $1.39 \pm 0.14$ [32] | $1.65 \pm 0.12$ | $1.38 \pm 0.1$ |
| Sapphire | $3.05 \pm 0.1$ [33] | $35 \pm 3.5$ [34] | $3.08 \pm 0.2$ | $37 \pm 2.6$ |
| Silicon | $1.665 \pm 0.05$ [35] | $142 \pm 2.8$ [36] | $1.665 \pm 0.09$ | $142 \pm 8$ |
| Diamond | $1.83 \pm 0.08$ [37] | $2000 \pm 200$ [38] | $1.93 \pm 0.077$ | $1890 \pm 173$ |

## 4. CONCLUSIONS

In summary, the SPS technique presented here marks a significant leap in thermal characterization, particularly for low-conductivity materials. Through simultaneous measurements of thermal conductivity and heat capacity with sub-millimeter-scale spatial resolution, the SPS method overcomes limitations associated with TDTR and FDTR, extending the measurable thermal conductivities down to $0.1~\mathrm{W/(m \cdot K)}$. The method's versatility is highlighted by its ability to comprehensively characterize both highly and lowly conductive materials. Validated against literature values, the SPS technique demonstrates remarkable reliability, with a measurement uncertainty below 10% across the entire thermal conductivity range from 0.1 to $2000~\mathrm{W/(m \cdot K)}$. This outstanding accuracy positions the SPS method as a robust tool for researchers and engineers seeking precise thermal property data for diverse materials.

## AUTHOR STATEMENT

All authors agreed to submit the manuscript to International Journal of Heat and Mass Transfer. The article is the authors' original work, hasn't received prior publication and isn't under consideration for publication elsewhere.

## DATA AVAILABILITY

The data that support the findings of this study are available from the corresponding author upon reasonable request.

## DECLARATION OF COMPETING INTEREST

The authors declare that they have no known competing financial interests or personal relationships that could have appeared to influence the work reported in this paper.

## ACKNOWLEDGMENT

This work is supported by the National Natural Science Foundation of China (NSFC) through Grant No. 52376058.